\documentclass[showpacs,groupedaddress,letterpaper,10pt,twocolumn,pra]{revtex4}
\usepackage{graphicx}

\input{tcilatex}

\begin{document}

\title{Entanglement generation in double-$\Lambda $ system}
\author{Ling Zhou, Yong Hong Ma, Xin Yu Zhao}
\affiliation{School of physics and optoelectronic technology, Dalian University of
Technology, Dalian 116024, P.R.China}

\begin{abstract}
In this paper, we study the generation of entanglement in a double-$\Lambda $
system. \ Employing standard method of laser theory, we deduce the dynamic
evolution equation of the two-mode field. We analyze the available
entanglement criterion for double-$\Lambda $ system and the condition of
entanglement existence. Our results show that under proper parameters, the
two-mode field can entangled and amlified.
\end{abstract}

\maketitle

\section{\protect\bigskip Introduction}

Continuous variables entanglement (CVE), as entanglement resource, has
attracted lots of attention because CVE not only has advantages in
quantum-information science \cite{1} but also can be prepared
unconditionally, whereas the preparation of discrete entanglement usually
relies on an event selection via coincidence measurements. Conventionally,
continuous variables entanglement has been produced by nondegenerate
parametric down- conversion (NPD) \cite{pan}. In order to improve the
strength of the NPD, engineering the NPD Hamiltonian within cavity QED has
also attracted much attention \cite{strength,zhou,guzman}. Besides
parametric down- conversion \cite{pan,zhang,simon1},\ Xiong et. al. \cite%
{han} had shown that two-photon correlated spontaneous emission laser can
work as a continuous variables entanglement producer and amplifier, which
open a new attracting research domain. And then, a number of different
schemes have been proposed [9-14]. Different from the gain medium atoms in
[8-14], Ref. \cite{LU} has studied a single-molecular-magnets system to
produce CVE where physics process is similar to \cite{kiffner}. All of these
works deal with the similar physics process where both of the two mode will
be created (annihilated) a photon in one loop respectively ( similar to
down-conversion system ).

In this paper, we proposed a scheme to generate CVE where the one mode is
created a photon and the other is annihilated, which is different from
[8-16]. The system consists of atoms in double-$\Lambda $ configuration
interacting with two modes cavity fields. The atoms are driven into a
coherent state of the upper two levels by two classical field. We obtain the
master equation of the two mode fields. Through analysis of entanglement, we
find that the criterion proposed in \cite{zubairy} can be used to judge
entanglement. We show that in double-$\Lambda $ system, entanglement exist
on the condition that the two-mode quantum field is tuned away from the
atomic transition, and the initial field is in a quantum state. Our study is
helpful to understand the entanglement charateristic within a system where
quantum field is in ''$V$'' configuration.

\section{The model and theory calculation}

We consider a system of atoms in double-$\Lambda $ configuration shown in
Fig.1. Two cavity fields interact with atomic transition $|a\rangle
\leftrightarrow |c\rangle $ and $|b\rangle \leftrightarrow |c\rangle $ with
detuning $\Delta _{a}$ and $\Delta _{b},$ respectively. The two classical
pumping fields with Rabi frequency $\Omega _{2}$ and $\Omega _{1}$ drive the
atomic level between $|a\rangle \leftrightarrow |d\rangle $ and $|b\rangle
\leftrightarrow |d\rangle $ with detuning $\Delta _{1}$ and $\Delta _{2}$
respectively. Our double-$\Lambda $ system can be sodium atoms in a vapor
cell \cite{EA} where the lower states are the two hyperfine levels 
\TEXTsymbol{\vert}$F=1\rangle $ and $F=2\rangle $ of 3$^{2}S_{1/2}$, and the
upper state are \TEXTsymbol{\vert}$F=1\rangle $ and $F=2\rangle $ of 3$%
^{2}P_{1/2}$. The double-$\Lambda $ system also can be atomic Pb vapor \cite%
{harris}. The phase-dependent electromagnetically induced transparency \cite%
{EA} and efficient nonlinear frequency conversion \cite{harris} have been
investigated experimentally in double-$\Lambda $ system. Ref. \cite{chong}
studied dark-state polaritons in double-$\Lambda $ system. Here, we are
interested in producing two-mode entangled laser via the double-$\Lambda $
system. In interaction picture, the Hamiltonian of the system can be written
as

\begin{eqnarray}
H_{0} &=&\nu _{1}a_{1}^{\dagger }a_{1}+\nu _{2}a_{2}^{\dagger }a_{2} \\
&&+\omega _{a}|a\rangle \langle a|+\omega _{b}|b\rangle \langle b|+\omega
_{c}|c\rangle \langle c|+\omega _{d}|d\rangle \langle d|  \nonumber
\end{eqnarray}%
\begin{figure}[tbph]
\includegraphics*[width=80mm, height=50mm]{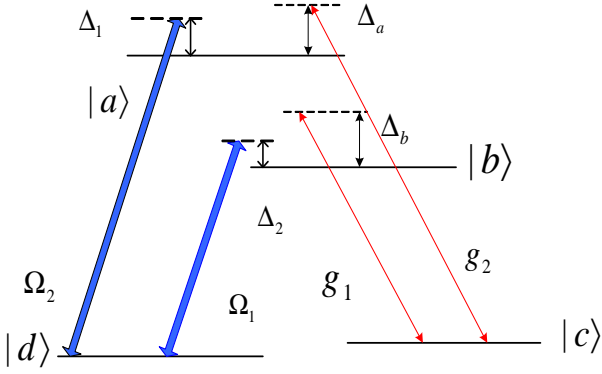}
\caption{The level configuration of atoms. Two cavity modes interact with
atomic transition $|b\rangle \leftrightarrow |c\rangle $ and $|a\rangle
\leftrightarrow |c\rangle $ with detuning $\Delta _{b}$ and $\Delta _{a}$
respectively while the two classical fields drive the atomic level between $%
|b\rangle \leftrightarrow |d\rangle $ and $|a\rangle \leftrightarrow
|d\rangle $ with detuning $\Delta _{2}$ and $\Delta _{1}$. For simplicity,
we assume the spontaneous-emission rate of four level are the same. }
\end{figure}
\begin{eqnarray}
H_{1} &=&g_{1}a_{1}|b\rangle \langle c|+g_{2}a_{2}|a\rangle \langle c| \\
&&+\Omega _{1}|b\rangle \langle d|e^{-i\omega _{1}t}+\Omega _{2}|a\rangle
\langle d|e^{-i\omega _{2}t}+H.c.  \nonumber
\end{eqnarray}%
We hope that the Hamiltonian do not contain time $t$ so as to simplify the
density matrix deduction of the field. In order to do that, we assume that
the classical fields detuning $\Delta _{1}=$ $\Delta a-\Delta $ and $\Delta
_{2}=\Delta _{b}$ $-\Delta $. Now we goes into a frame by performing a
unitary transformation $U=exp\{i[H_{0}+\Delta _{a}|a\rangle \langle
a|+\Delta _{b}|b\rangle \langle b|+\Delta |d\rangle \langle d|]t\}$. In the
new frame, the Hamiltonian is as

\begin{eqnarray}
H_{1} &=&-\Delta a|a\rangle \langle a|-\Delta b|b\rangle \langle b|-\Delta
|d\rangle \langle d|  \nonumber \\
&&+[g_{1}a_{1}|b\rangle \langle c|+g_{2}a_{2}|a\rangle \langle c| \\
&&+\Omega _{1}|b\rangle \langle d|e^{-i\omega _{1}t}+\Omega _{2}|a\rangle
\langle d|e^{-i\omega _{2}t}+H.c.].  \nonumber
\end{eqnarray}%
In order to see the entanglement of the two-mode field, we need to obtain
the equation of motion of the two-mode field. Using the standard procedure
in laser theory developed by Scully and Zubairy\cite{scully,scully2,walls},
we obtain the following master equation governing the dynamics of the
two-mode cavity fields as%
\begin{eqnarray}
\dot{\rho} &=&-\kappa _{1}(a_{1}^{\dagger }a_{1}\rho -a_{1}\rho
a_{1}^{\dagger })-\kappa _{2}(a_{2}^{\dagger }a_{2}\rho -a_{2}\rho
a_{2}^{\dagger })  \label{A14} \\
&&-\alpha _{1}(\rho a_{1}a_{1}^{\dagger }-a_{1}^{\dagger }\rho a_{1})-\alpha
_{2}(\rho a_{2}a_{2}^{\dagger }-a_{2}^{\dagger }\rho a_{2})  \nonumber \\
&&-\alpha _{12}(\rho a_{2}a_{1}^{\dagger }-a_{1}^{\dagger }\rho
a_{2})-\alpha _{21}(\rho a_{1}a_{2}^{\dagger }-a_{2}^{\dagger }\rho
a_{1})+h.c..  \nonumber
\end{eqnarray}%
We can see that the master equation has the term $\rho a_{2}a_{1}^{\dagger
}-a_{1}^{\dagger }\rho a_{2}$ which means that the one mode is created a
photon and the other mode is annihilated a photon. The detail deduction of
the equation is given in appendix A. In Eq.(4), we have include the loss of
the two-mode cavity with loss rate $\kappa _{1}$ and $\kappa _{2}.$ The
coefficients are 
\begin{eqnarray}
\alpha _{1} &=&\frac{g_{1}^{2}}{D}[A_{11}L_{bb}+A_{21}L_{ab}+A_{31}L_{db}],
\\
\alpha _{2} &=&\frac{g_{2}^{2}}{D}[A_{12}L_{ba}+A_{22}L_{aa}+A_{32}L_{da}], 
\nonumber \\
\alpha _{12} &=&\frac{g_{1}g_{2}}{D}[A_{11}L_{ba}+A_{21}L_{aa}+A_{31}L_{da}],
\nonumber \\
\alpha _{21} &=&\frac{g_{1}g_{2}}{D}[A_{12}L_{bb}+A_{22}L_{ab}+A_{32}L_{db}],
\nonumber
\end{eqnarray}%
where $D=(\gamma -i\Delta _{a})(\gamma -i\Delta _{b})[\gamma -i(\Delta +%
\frac{\Delta _{a}}{2}+\frac{\Delta _{b}}{2}]+\Omega _{1}^{2}(\gamma -i\Delta
_{a})+\Omega _{2}^{2}(\gamma -i\Delta _{b})$, and%
\begin{eqnarray}
L_{aa} &=&\frac{-i\Omega _{2}}{\gamma }y_{3},L_{bb}=\frac{-i\Omega _{1}}{%
\gamma }y_{2}, \\
L_{ab} &=&\frac{\gamma +i(\Delta _{2}-\Delta _{1})}{2\gamma }y_{1}-\frac{%
i\Omega _{2}}{2\gamma }y_{2}-\frac{-i\Omega _{1}}{2\gamma }y_{3},  \nonumber
\\
L_{db} &=&\frac{\gamma -i\Delta _{2}}{2\gamma }y_{2}-\frac{i\Omega _{2}}{%
2\gamma }y_{1},  \nonumber \\
L_{da} &=&\frac{\gamma -i\Delta _{1}}{2\gamma }y_{3}+\frac{i\Omega _{1}}{%
2\gamma }y_{1},  \nonumber
\end{eqnarray}%
and \ 
\begin{eqnarray}
A_{11} &=&(\gamma -i\Delta _{a})[\gamma -i(\Delta +\frac{\Delta _{a}}{2}+%
\frac{\Delta _{b}}{2}], \\
A_{12} &=&A_{21}=-\Omega _{1}\Omega _{2},A_{31}=-i\Omega _{1}(\gamma
-i\Delta _{a}),  \nonumber \\
A_{22} &=&(\gamma -i\Delta _{b})[\gamma -i(\Delta +\frac{\Delta _{a}}{2}+%
\frac{\Delta _{b}}{2}],  \nonumber \\
A_{32} &=&-i\Omega _{2}(\gamma -i\Delta _{b})  \nonumber
\end{eqnarray}%
with%
\begin{eqnarray}
y_{2} &=&\frac{2ir_{in}s(a_{1}\Omega _{2}-b\Omega _{1})}{a_{1}a_{2}-b^{2}}%
,y_{3}=\frac{2ir_{in}s(a_{2}\Omega _{1}-b\Omega _{2})}{a_{1}a_{2}-b^{2}} \\
y_{1} &=&\frac{\Omega _{2}(\Delta _{1}-2\Delta _{2})}{s}y_{2}+\frac{\Omega
_{1}(2\Delta _{1}-\Delta _{2})}{s}y_{3},  \nonumber
\end{eqnarray}%
in which%
\begin{eqnarray}
s &=&\gamma ^{2}+\Omega _{1}^{2}+\Omega _{2}^{2}+(\Delta _{2}-\Delta
_{1})^{2}, \\
a_{1} &=&M_{1}s-\Omega _{2}^{2}(2\Delta _{2}-\Delta _{1})^{2},  \nonumber \\
a_{2} &=&M_{2}s-\Omega _{1}^{2}(2\Delta _{1}-\Delta _{2})^{2},  \nonumber \\
b &=&\Omega _{1}\Omega _{2}[3s-(\Delta _{1}-2\Delta _{2})(2\Delta
_{1}-\Delta _{2})]  \nonumber \\
M_{1} &=&\gamma ^{2}+4\Omega _{1}^{2}+\Omega _{2}^{2}+\Delta _{2}^{2}, 
\nonumber \\
M_{2} &=&\gamma ^{2}+4\Omega _{2}^{2}+\Omega _{1}^{2}+\Delta _{1}^{2}. 
\nonumber
\end{eqnarray}

Although our four-level atom is similar to \cite{kiffner, LU}, the physical
process of the two-mode quantum fields is different because the two quantum
fields work in different atomic level. In [8-15], both of the two mode will
be created or annihilated a photon in one loop. So the master equation is of
the form $\rho a_{2}^{\dagger }a_{1}^{\dagger }-a_{1}^{\dagger }\rho
a_{2}^{\dagger }$ ($\rho a_{2}a_{1}-a_{1}\rho a_{2}$). In our system, the
two quantum fields are in a ''V'' form levels if we do not see the two
classical pumping fields. The simplified ''V'' form levels is similar to
''Hanle effect'' laser \cite{hanle} where the master equation is with the
term $\rho a_{2}a_{1}^{\dagger }-a_{1}^{\dagger }\rho a_{2}$. In our system,
the two classical fields make the atoms with the coherence of the two
up-level $|a\rangle $ and $|b\rangle $ [see (A10)]. When the spontaneous
emissions from $|a\rangle $ and $|b\rangle $ to $|c\rangle $ take place,
entangled photons will be produced.

\section{Entanglement criterion choice and the discussion of the
entanglement condition}

How to determine the entanglement is a key problem. In Ref.[8-15], employing
the criterion $(\Delta u)^{2}+(\Delta v)^{2}<2$\cite{duan}, a inequality of
the sum of the quantum fluctuations of two operators $u$ and $v$ for some
entangled state, they find the entanglement between the two mode fields.
However, the criterion inequality of the sum of the quantum fluctuations can
not be applied to measure coherent state\cite{shchu}. Although the
entanglement criterion on measure continuous variable have been
developed[22-25], we still can not find a criterion to judge all kind of
continuous variable entanglement. In order to make clear the kind of
entanglement existing in our model, we now discuss the analytic solution in
our system so as to choose a appropriate entanglement criterion as well as
to know the condition of entanglement.

Now we analyze the entanglement condition . If $g_{1}=g_{2}$, $\Omega
_{1}=\Omega _{2}$, and $\Delta =\Delta _{a}=\Delta _{b}\gg $ $\Omega
_{1},\Omega _{2},\gamma $, through Eq.(6) to (10), one can obtain the
relation $\alpha _{1}=\alpha _{2}=\alpha _{12}=\alpha _{21}=i\alpha $ ($%
\alpha $ is a real number). Usually, the loss of the cavity do not change
the entanglement structure of the state. It just destroy or sometimes
enhance the entanglement a little. So, in our choice entanglement criterion,
we omit the loss of the cavity. Therefore, the master equation of our system
Eq.(4) can be simplified as 
\begin{equation}
\dot{\rho}=i\alpha \lbrack a_{1}a_{1}^{\dagger }+a_{2}a_{2}^{\dagger
}+a_{2}a_{1}^{\dagger }+a_{1}a_{2}^{\dagger },\rho ].
\end{equation}%
The effective Hamiltonian $H_{eq}=-\alpha (a_{1}a_{1}^{\dagger
}+a_{2}a_{2}^{\dagger }+a_{2}a_{1}^{\dagger }+a_{1}a_{2}^{\dagger })$. Due
to $[a_{1}a_{1}^{\dagger }+a_{2}a_{2}^{\dagger }$,$a_{2}a_{1}^{\dagger
}+a_{1}a_{2}^{\dagger }]=0$,in interaction picture $H_{eqI}=$ $-\alpha
(a_{2}a_{1}^{\dagger }+a_{1}a_{2}^{\dagger })$. One can easy check that the
system state, evolved by $H_{eqI}=$ $-\alpha (a_{2}a_{1}^{\dagger
}+a_{1}a_{2}^{\dagger }),$ never meet with the criterion $(\Delta
u)^{2}+(\Delta v)^{2}<r^{2}+\frac{1}{r^{2}}$ for the initial field number $%
|n_{1,}n_{2}\rangle $. We recognize the field Hamiltonian is the generator
of the $SU(2)$ coherent state \cite{gerry}. The evolution of the state $%
|\Psi (0)\rangle $ is%
\[
|\Psi (t)\rangle =e^{-iH_{eqI}t}|\Psi (0)\rangle =e^{x_{+}K_{+}}e^{K_{0}\ln {%
x_{0}}}e^{x_{-}K_{-}}|\Psi (0)\rangle , 
\]%
where $K_{+}=a_{1}^{\dagger }a_{2}$, $K_{-}=a_{1}a_{2}^{\dagger }$. These
operators satisfy the $SU(2)$ commutation relations, i.e., $%
[K_{-},K_{+}]=-2K_{0}$, $[K_{0},K_{+}]=K_{+}$, $[K_{0},K_{-}]=-K_{-}$, with $%
K_{0}=\frac{1}{2}(a_{1}^{\dagger }a_{1}-a_{2}^{\dagger }a_{2})$; and in
which 
\begin{eqnarray*}
x_{0} &=&\{\cosh i\alpha t\}^{-\frac{1}{2}}, \\
x_{+} &=&x_{-}=\tanh i\alpha t.
\end{eqnarray*}%
If the initial field state is two-mode Fock state $|0,N\rangle $, the
evolution of the state is 
\begin{equation}
|\Psi (t)\rangle =(\cos \alpha t)^{N/2}\sum_{n=0}^{N}\left( 
\begin{array}{c}
N \\ 
n%
\end{array}%
\right) ^{1/2}(i\tan \alpha t)^{n}|n,N-n\rangle .
\end{equation}%
From the entanglement definition of pure state, we know that the state $%
|\Psi (t)\rangle $ is a entangled one.

Unfortunately, the Hamiltonian $H_{eqI}$ can not entangle initial coherent
state, because the evolution of the system as 
\begin{equation}
|\Psi (t)\rangle =e^{x_{+}K_{+}}e^{K_{0}\ln {x_{0}}}e^{x_{-}K_{-}}|\beta
_{1},\beta _{2}\rangle =|\tilde{\beta}_{1},\tilde{\beta}_{2}\rangle ,
\end{equation}%
with%
\begin{eqnarray*}
\tilde{\beta}_{1} &=&\beta _{1}\cos \alpha t+i\beta _{2}\sin \alpha t, \\
\tilde{\beta}_{2} &=&\beta _{2}\cos \alpha t+i\beta _{1}\sin \alpha t.
\end{eqnarray*}%
So, it is not entangled.

The two-mode $SU(2)$ cat state is sub-Poissonian distribution. We recall the
criterion, proposed by Hillery and Zubairy\cite{zubairy} can be used for
non- Gaussionian state. The criterion say that if 
\begin{equation}
\langle N_{1}N_{2}\rangle <|\langle a_{1}a_{2}^{\dagger }\rangle |^{2}
\end{equation}%
the two-mode field is entangled. If the field initially is in number state $%
|n_{1},n_{2}\rangle $, using the differential equation Eq.(B1-B13)(let $%
\kappa =0$ and $\alpha _{1}=\alpha _{2}=\alpha _{12}=\alpha _{21}=i\alpha $%
), we finally obtain $\ $%
\begin{equation}
\langle N_{1}N_{2}\rangle -|\langle a_{1}a_{2}^{\dagger }\rangle
|^{2}=n_{1}n_{2}-\frac{1}{4}(n_{1}+n_{2}+2n_{1}n_{2})\sin ^{2}2\alpha t.
\end{equation}%
The maximum value of $\sin ^{2}2\alpha t$ is 1; therefore if 
\begin{equation}
2n_{1}n_{2}<n_{1}+n_{2},
\end{equation}%
the two mode field will be entangled. Because $n_{1}$ and $n_{2}$ are
integer, in order to meet with $2n_{1}n_{2}<n_{1}+n_{2}$, the number $n_{1}$
and $n_{2}$ should be not equal. If either $n_{1}$ or $n_{2}$ is zero (the
state is standard $SU(2)$ coherent state), we can see that $\langle
N_{a}N_{b}\rangle -|\langle ab^{\dagger }\rangle |^{2}$ is always less than
zero; thus we say the state is entangled. Therefore, the criterion Eq.(12)
can be used for judge entanglement within our system.

However, for resonant case ($\Delta _{b}=\Delta _{a}=\Delta =0$), if $\gamma
_{b}=\gamma _{a}$ , $g_{1}=g_{2}$ and $\Omega _{1}=\Omega _{2}$, the
coefficients $\alpha _{1}=\alpha _{2}=\alpha _{12}=\alpha _{21}=\beta $
(real number). For the initial state $|n_{1},n_{2}\rangle $, after
complicated calculation employing Eqs. B1-B13 for $\alpha _{1}=\alpha
_{2}=\alpha _{12}=\alpha _{21}=\beta $, we have%
\begin{eqnarray}
\langle N_{1}N_{2}\rangle -|\langle a_{1}a_{2}^{\dagger }\rangle |^{2} &=&%
\frac{n_{1}+n_{2}}{16}(3e^{8\beta t}+2e^{4\beta t}-5) \\
&&+\frac{n_{1}n_{2}}{8}(1+6e^{4\beta t}+e^{8\beta t})  \nonumber \\
&&+\frac{1}{4}(1+e^{8\beta t}-2e^{4\beta t})\succeq 0  \nonumber
\end{eqnarray}

If $n_{1}$ or $n_{2}$ is zero, $\langle N_{1}N_{2}\rangle -|\langle
a_{1}a_{2}^{\dagger }\rangle |^{2}$ equal to zero at initial time. Except
that the $\langle N_{1}N_{2}\rangle -|\langle a_{1}a_{2}^{\dagger }\rangle
|^{2}$ is larger than zero. It is obvious that we can not obtain
entanglement in resonant case. This conclusion is consistent with the work
in Ref.\cite{manzoor}, where author show that for two-level quantum beat
laser, entanglement can be created only when the strong driving field should
be tuned away from the atomic transition. 
\begin{figure}[h]
\includegraphics*[width=75mm, height=50mm]{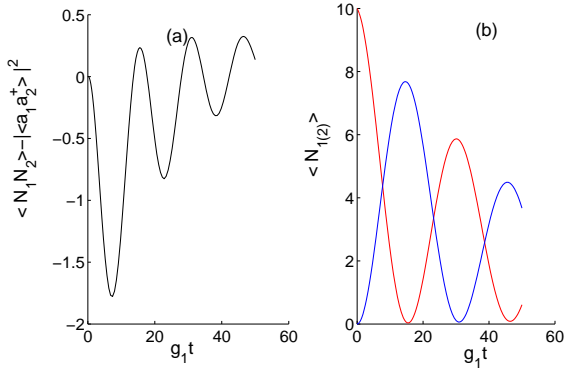}
\caption{(a): The time evolution of the entanglement.(b): The time evolution
of the two-mode fields where red line is for $N_{1}$ and blue line is for $%
N_{2.}$ Initially, the atom is in number state $|10,0\rangle $ The
parameters are $g_{1}=g_{2}=1;$ $\Delta _{a}=\Delta _{b}=50$, $\Delta =4$, $%
r_{in}=20$, $\protect\gamma =1$, $\protect\kappa _{1}=\protect\kappa %
_{2}=0.010$. $\Omega _{2}=\Omega _{1}=5$.}
\end{figure}

\section{\protect\bigskip The entanglement of the cavity field}

In above section, we discuss a special case so as to choose entanglement
criterion and make clear the condition of entanglement. Although above
analysis is for pure state (approximation of master equation Eq.(4)), But
the criterion $\langle N_{1}N_{2}\rangle <|\langle a_{1}a_{2}^{\dagger
}\rangle |^{2}$ should be available in judging entanglement for general
case. Now, considering the loss of the cavity and the decay of the atomic
levels, we numerical solve the differential Eqs. (B1) to (B13) and plot the
entanglement criterion $\langle N_{1}N_{2}\rangle -|\langle
a_{1}a_{2}^{\dagger }\rangle |^{2}$ and the $N_{1}(\langle a_{1}^{\dagger
}a_{1}\rangle )$,$N_{2}$($\langle a_{2}^{\dagger }a_{2}\rangle $).

In Fig.2, we plot the case that the initial field state is \ a number state $%
|10,0\rangle $ where $\Delta _{b}=\Delta _{a}\gg \gamma _{b}=\gamma _{a}$
which means that the classical field resonantly drive the atom ($\Delta
_{1}=\Delta _{2}=0)$ and the quantum field interact with the atoms with
equal detunings. We see that due to the loss of the cavity, the entanglement
gradually disappear and photon number of the two-mode field also decrease
under large detuning case. Of course, if the cavity is ideal, one will
observe the entanglement oscillation. 
\begin{figure}[tbph]
\includegraphics*[width=80mm, height=50mm]{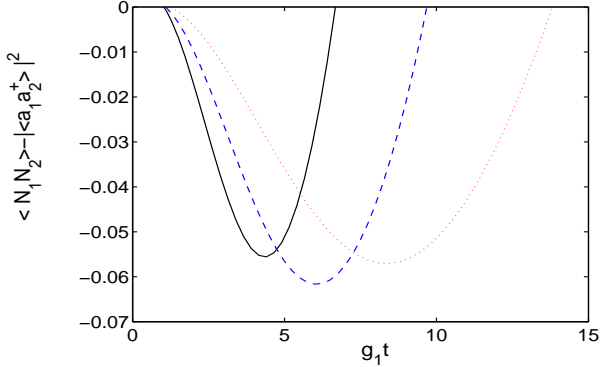}
\caption{The time evolution of entanglement. Initially, the atom is in
number state $|1,0\rangle $ The parameters are $g_{1}=g_{2}=1;$ $\Delta
_{a}=50,\Delta _{b}=20$, $\Delta =10$, $r_{in}=20$, $\protect\gamma =1$, $%
\protect\kappa _{1}=\protect\kappa _{2}=0.01$. $\Omega _{2}=\Omega
_{1}=4,5,6 $ for dotted, dashed and solid line, respectively.}
\end{figure}

However, with the same detuning $\Delta _{b}=\Delta _{a}$ (when $%
g_{1}=g_{2}) $, we can not have amplified entangled laser shown in Fig.2.
The quantum fields are in ''V'' form. If $\Delta _{b}=\Delta _{a}$, the
photon number in two mode only oscillate because of the symmetry. In our
numerical simulation, we find that in order to have amplified entangled
laser, $\Delta _{a}$ and $\Delta _{b}$ should be different. For initial
field state in number state $|1,0\rangle $, we plot entanglement and average
photon numbers in Fig.3 and 4 for several values of $\Omega _{1}(\Omega
_{2}) $. One can see clearly that entanglement can be obtained without
preparation atomic coherence before (here, atoms are injected in state $%
|d\rangle $). But the photon number in two mode has large difference. By
adjusting the values of $\Omega _{1}(\Omega _{2})$, we can adjust the time
region of entanglement. Because we inject the atom in atomic state $%
|d\rangle $, it will need time to evolve into a coherence among the atomic
level $|a\rangle $, $|b\rangle $ and $|d\rangle $. So, we have no
entanglement during a initial short time . With large value of $\Omega
_{1}(\Omega _{2})$, the atoms will acquire their coherence quickly so that
the entanglement appear quickly. However, with large value of $\Omega
_{1}(\Omega _{2})$, the photon number also will be \ amplified quickly shown
in Fig. 4. As in our analytic calculation, we have known that the photon
number in two mode differ (Eq.(15)). Here, in order to amplify the photon
number, the photon number not only should have difference but also can not
put up with very large photon number. With the increasing of photon number,
the entanglement disappear. But the disentanglement is not resulted from
loss of the cavity because we find even for $\kappa =0$, entanglement also
disappear. We conclude that the disentanglement result from the increase of
photon number rather than from the loss of the cavity. As it is pointed out
in Ref.[21],in the high-gain limit the condition in Eq.(9) is no longer able
to detect whether there is entanglement in the state.

\begin{figure}[tbph]
\includegraphics*[width=80mm, height=50mm]{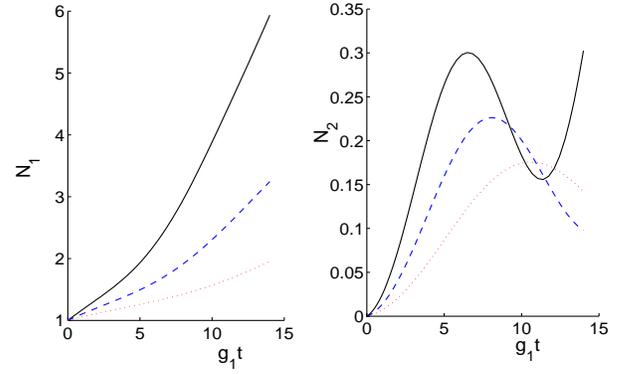}
\caption{The time evolution of average photon number. }
\end{figure}

Now, we show another function of the classical fields,i.e., the ability to
overcome the loss of the cavity which is shown in Fig.4. Let us compare
dotted line and solid line. The two lines correspond to the loss rate of the
cavity $\kappa _{1}=\kappa _{2}=0.01$ and $0.1$, respectively; and all the
other parameters are the same. Due to the increasing of\ $\kappa _{1}$($%
\kappa _{2}$), the values of $\langle N_{1}N_{2}\rangle -|\langle
a_{1}a_{2}^{\dagger }\rangle |^{2}$ move up. If $\kappa _{1}$($\kappa _{2}$)
keep increasing, we will loss entanglement. However, with the help of
classical fields, we still can obtain entanglement even through $\kappa _{1}$%
($\kappa _{2}$) is large, which can be observed by comparing dashed line and
solid one. Although the loss rate $\kappa _{1}=\kappa _{2}=0.1$, through
increasing $\Omega _{1}(\Omega _{2})$ to $6$, we still can have
entanglement. Of course, because of the increasing of $\Omega _{1}(\Omega
_{2})$, the time region move left, which we have analyze it in Fig.3.

\begin{figure}[tbph]
\includegraphics*[width=80mm, height=50mm]{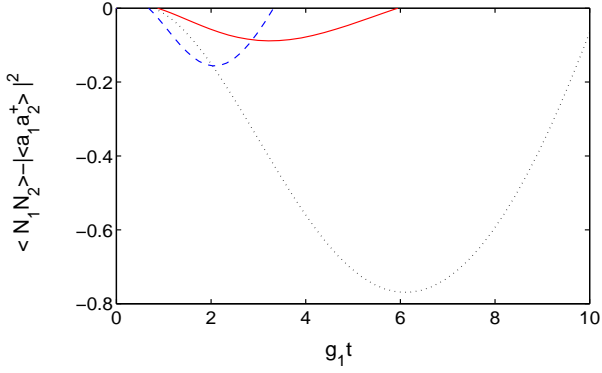}
\caption{ The time evolution of entanglement. Initially, the atom is in
number state $|10,0\rangle $ The parameters are $g_{1}=g_{2}=1;$ $\Delta
_{a}=50,\Delta _{b}=20$, $\Delta =10$, $r_{in}=30$, $\protect\gamma =1$.
Dotted line: $\protect\kappa _{1}=\protect\kappa _{2}=0.010$. $\Omega
_{2}=\Omega _{1}=4;$solid line: $\protect\kappa _{1}=\protect\kappa _{2}=0.1$%
, $\Omega _{2}=\Omega _{1}=4$; dashed line:$\protect\kappa _{1}=\protect%
\kappa _{2}=0.1$, $\Omega _{2}=\Omega _{1}=6.$}
\end{figure}

\section{\protect\bigskip \protect\bigskip Conclusion}

In conclusion, we have studied the generation of entanglement in a double-$%
\Lambda $ system. We derive the theory of this system and analyze the
available entanglement criterion for double-$\Lambda $ system. When the
atoms are injected in the ground state $|d\rangle $, the entangled laser can
be achieved under the condition of suitable parameters. Due to the classical
pumping field introduction, we do not need to prepare atomic coherence, and
the intensity of the quantum fields will be amplified. The classical pumping
can overcome the loss of the cavity. Our results show that the time for
which the two modes remain entangled depends upon the strength of the Rabi
frequency of the classical driving field.

Our results is helpful in understanding the entanglement characteristic when
the master equation contain the term $\rho a_{2}a_{1}^{\dagger
}-a_{1}^{\dagger }\rho a_{2}$ such as quantum beats laser and Hanle effect
laser system. Our studies is limited to the initial state $|1,0\rangle $.
One can research other initial field state. Our initial field should be easy
to obtain. Let excited two-level atom with transition frequency $\nu 1($ or $%
\nu 2)$ passing through the vacuum two-mode cavity, when we detect the
output atom in ground state, we will have the field state $|1,0\rangle $.

Acknowledgments: Authors thank Professor M. S. Zubairy and M. Ikram for
their critical reading. The project was supported by NSFC under Grant
No.10774020, and also supported by SRF for ROCS, SEM.\bigskip

\appendix

\section{Calculation details of density matrix of two-mode fields}

The classical fields will be treated to all orders in the Rabi frequency.
The transitions $|a\rangle $ - $|c\rangle $ and $|b\rangle $ - $|c\rangle $
are treated fully quantum mechanically but only up to second order in the
corresponding coupling constants. By partially tracing the global state of
Schr\"{o}dinger equation over the atomic variables, we have the formal
reduced fields%
\begin{equation}
\dot{\rho}_{f}=-i([H_{cb,}\rho _{bc}]+[H_{ca},\rho _{ac}]+[H_{ac},\rho
_{ca}]+[H_{bc},\rho _{cb}]).  \label{A1}
\end{equation}%
Eq.(\ref{A1}) reveals that we need to get the density matrix elements $\rho
_{ca},\rho _{bc,}$ etc.. Inserting the Hamiltonian Eq.(1) into A.1, from the
schr\"{o}dinger equation, we have 
\begin{eqnarray}
\dot{\rho}_{bc} &=&-(\gamma -i\Delta _{b})\rho _{bc}-i\Omega _{1}\rho _{dc}
\label{A2} \\
&&+(-ig_{1}a_{1}\rho _{cc}+ig_{1}\rho _{bb}a_{1}+ig_{2}\rho _{ba}a_{2}), 
\nonumber \\
\dot{\rho}_{ac} &=&-(\gamma -i\Delta _{a})\rho _{ac}-i\Omega _{2}\rho _{dc} 
\nonumber \\
&&+(ig_{1}\rho _{ab}a_{1}+ig_{2}\rho _{aa}a_{2}-ig_{2}a_{2}\rho _{cc}), 
\nonumber \\
\dot{\rho}_{dc} &=&-[\gamma -i\Delta ]\rho _{dc}-i\Omega _{1}\rho
_{bc}-i\Omega _{2}\rho _{ac}  \nonumber \\
&&+(ig_{1}\rho _{db}a_{1}+ig_{2}\rho _{da}a_{2}).  \nonumber
\end{eqnarray}%
In the last equations Eq.(\ref{A2}), we have consider the
spontaneous-emission of the atomic level. We rewrite it in a matrix form as%
\begin{equation}
\dot{\rho}=-M\rho +A  \label{A3}
\end{equation}%
where%
\begin{equation}
\rho =\left( 
\begin{array}{c}
\rho _{bc} \\ 
\rho _{ac} \\ 
\rho _{dc}%
\end{array}%
\right) ,  \label{A4}
\end{equation}%
\begin{equation}
M=\left( 
\begin{array}{ccc}
\gamma -i\Delta _{b} & 0 & i\Omega _{1} \\ 
0 & \gamma -i\Delta _{a} & i\Omega _{2} \\ 
i\Omega _{1} & i\Omega _{2} & \gamma -i\Delta%
\end{array}%
\right) ,  \label{A5}
\end{equation}%
\begin{equation}
A=\left( 
\begin{array}{c}
ig_{1}\rho _{bb}a_{1}+ig_{2}\rho _{ba}a_{2} \\ 
ig_{1}\rho _{ab}a_{1}+ig_{2}\rho _{aa}a_{2} \\ 
ig_{1}\rho _{db}a_{1}+ig_{2}\rho _{da}a_{2}%
\end{array}%
\right) .
\end{equation}%
When we write matrix $A$, we let $\rho _{cc}=0$ and will explain the reason
later. A solution of Eq.(\ref{A3}) which is a linear in the coupling
constant $g_{1(2)}$ can be obtained \cite{scully,scully2,walls}. Here we
only care for the matrix elements $\rho _{bc}$ and $\rho _{ac}$, so we just
write the solution of the two terms as%
\begin{eqnarray}
\rho _{bc} &=&\frac{i}{D}[(A_{11}\rho _{bb}^{0}+A_{21}\rho
_{ab}^{0}+A_{31}\rho _{db}^{0})g_{1}a_{1}  \nonumber \\
&&+(A_{11}\rho _{ba}^{0}+A_{21}\rho _{aa}^{0}+A_{31}\rho
_{da}^{0})g_{2}a_{2}],  \label{A7}
\end{eqnarray}%
\begin{eqnarray}
\rho _{ac} &=&\frac{i}{D}[(A_{12}\rho _{bb}^{0}+A_{22}\rho
_{ab}^{0}+A_{32}\rho _{db}^{0})g_{1}a_{1}  \nonumber \\
&&+(A_{12}\rho _{ba}^{0}+A_{22}\rho _{aa}^{0}+A_{32}\rho
_{da}^{0})g_{2}a_{2}]  \label{A8}
\end{eqnarray}%
with%
\begin{eqnarray}
A_{11} &=&(\gamma -i\Delta _{a})(\gamma -i\Delta )+\Omega _{2}^{2}, \\
A_{12} &=&A_{21}=-\Omega _{1}\Omega _{2},A_{31}=-i\Omega _{1}(\gamma
-i\Delta _{a}),  \nonumber \\
A_{22} &=&(\gamma -i\Delta _{b})(\gamma -i\Delta )+\Omega _{1}^{2}, 
\nonumber \\
A_{32} &=&-i\Omega _{2}(\gamma -i\Delta _{b}),  \nonumber
\end{eqnarray}%
where $D=(\gamma -i\Delta _{a})(\gamma -i\Delta _{b})(\gamma -i\Delta
)+\Omega _{1}^{2}(\gamma -i\Delta _{a})+\Omega _{2}^{2}(\gamma -i\Delta
_{b}) $ in Eqs.(\ref{A7})(\ref{A8}). As a approximation, the density matrix
elements in right side of Eqs.(\ref{A7})(\ref{A8}) such as $\rho _{bb}^{0}$, 
$\rho _{ba}^{0}$, etc. will be determined by steady state of classical
fields. In other words, the density matrix elements $\rho _{bb}^{0}$, $\rho
_{ba}^{0}$, etc. of classical fields, as a zero order approximation, are
substituted into right side of Eqs.(\ref{A7}) and (\ref{A8}) . And then, we
can obtain a first order approximation of density matrix elements $\rho
_{ca} $ , $\rho _{ab}$ in terms of couplings $g_{1}(g_{2)}$.

Now, we just consider classical fields to determine the zero order
approximation of the density matrix elements $\rho _{bb}^{0}$, $\rho
_{ba}^{0}$, etc.. The differential equations of density matrix elements only
with classical fields and atomic decay are 
\begin{eqnarray}
\dot{\rho}_{bb}^{0} &=&-\gamma \rho _{bb}^{0}-i\Omega _{1}(\rho
_{db}^{0}-\rho _{bd}^{0}),  \label{A9} \\
\dot{\rho}_{aa}^{0} &=&-\gamma \rho _{aa}^{0}-i\Omega _{2}(\rho
_{da}^{0}-\rho _{ad}^{0}),  \nonumber \\
\dot{\rho}_{ba}^{0} &=&-[\gamma -i(\Delta _{2}-\Delta _{1})]\rho
_{ba}^{0}+i\Omega _{2}\rho _{bd}^{0}-i\Omega _{1}\rho _{da}^{0},  \nonumber
\\
\dot{\rho}_{da}^{0} &=&-[\gamma +i\Delta _{1})]\rho _{da}^{0}-i\Omega
_{2}(\rho _{aa}^{0}-\rho _{dd}^{0})-i\Omega _{1}\rho _{ba}^{0},  \nonumber \\
\dot{\rho}_{db}^{0} &=&-[\gamma +i\Delta _{2})]\rho _{db}^{0}-i\Omega
_{2}\rho _{ab}^{0}-i\Omega _{1}(\rho _{bb}^{0}-\rho _{dd}^{0}),  \nonumber \\
\dot{\rho}_{dd}^{0} &=&-\gamma \rho _{db}^{0}-i\Omega _{1}(\rho
_{bd}^{0}-\rho _{db}^{0})-i\Omega _{2}(\rho _{ad}^{0}-\rho
_{da}^{0})+r_{in}\rho  \nonumber
\end{eqnarray}%
with $\Delta _{1}=\Delta _{a}-\Delta ,\Delta _{2}=\Delta _{b}-\Delta $. For $%
\dot{\rho}_{cc}$, we have $\dot{\rho}_{cc}^{0}=-\gamma \rho _{cc}^{0}$. The
steady state solution $\rho _{cc}^{0}=0$ (It is the reason why we let $\rho
_{cc}=0$ in Eq.(A6)). Substituting the steady state solution of (A10) into
(A7) and (A8), we obtain $\rho _{bc}$ , $\rho _{ac}$. And then, inserting $%
\rho _{bc}$ and $\rho _{ac}$ back into (A1), one can have the master
equation Eq.(4) with coefficients Eq.(5) to (9).

\section{Calculation details of density matrix of two-mode fields}

In order to numerical calculate the entanglement criterion $\langle
N_{1}N_{2}\rangle -|\langle a_{1}a_{2}^{\dagger }\rangle |^{2}$ and the $%
N_{1}(N_{2})$, we need to deduce a series differential equations from master
equations (4) which are listed below.

\begin{eqnarray}
\frac{d\langle a_{1}^{\dagger }a_{1}\rangle }{dt} &=&(\alpha _{1}+\alpha
_{1}^{\ast }-2\kappa _{1})\langle a_{1}^{\dagger }a_{1}\rangle  \\
&&+\alpha _{12}^{\ast }\langle a_{2}^{\dagger }a_{1}\rangle +\alpha
_{12}\langle a_{2}a_{1}^{\dagger }\rangle +\alpha _{1}+\alpha _{1}^{\ast }, 
\nonumber \\
\frac{d\langle a_{1}a_{2}^{\dagger }\rangle }{dt} &=&(\alpha _{1}+\alpha
_{2}^{\ast }-\kappa _{1}-\kappa _{2})\langle a_{1}a_{2}^{\dagger }\rangle  \\
&&+\alpha _{21}^{\ast }\langle a_{1}^{\dagger }a_{1}\rangle +\alpha
_{12}\langle a_{2}^{\dagger }a_{2}\rangle +\alpha _{12}+\alpha _{21}^{\ast },
\nonumber
\end{eqnarray}%
\begin{eqnarray}
\frac{d\langle a_{2}^{\dagger }a_{2}a_{2}^{+}a_{1}\rangle }{dt} &=&(\alpha
_{1}+\alpha _{2}+2\alpha _{2}^{\ast }-\kappa _{1}-3\kappa _{2})\langle
a_{2}^{\dagger }a_{2}a_{2}^{+}a_{1}\rangle   \nonumber \\
&&+(\alpha _{2}+\alpha _{2}^{\ast }+2\kappa _{2})\langle
a_{2}^{+}a_{1}\rangle +2\alpha _{21}^{\ast }\langle a_{2}^{\dagger
}a_{2}a_{1}^{+}a_{1}\rangle   \nonumber \\
&&+2\alpha _{21}^{\ast }\langle a_{2}^{\dagger }a_{2}\rangle +\alpha
_{21}\langle a_{2}^{\dagger 2}a_{1}^{2}\rangle +\alpha _{12}\langle
a_{2}a_{2}^{\dagger }a_{2}a_{2}^{+}\rangle   \nonumber \\
&&+\alpha _{21}^{\ast }\langle a_{1}^{+}a_{1}\rangle +\alpha _{21}^{\ast },
\end{eqnarray}%
\begin{eqnarray}
\frac{d\langle a_{2}^{2}a_{1}^{+2}\rangle }{dt} &=&2(\alpha _{1}^{\ast
}+\alpha _{2}-\kappa _{2}-\kappa _{1})\langle a_{2}^{2}a_{1}^{+2}\rangle  \\
&&+2\alpha _{21}\langle a_{1}^{\dagger }a_{1}a_{1}^{+}a_{2}\rangle +2(\alpha
_{21}+\alpha _{12}^{\ast })\langle a_{1}^{+}a_{2}\rangle   \nonumber \\
&&+2\alpha _{12}^{\ast }\langle a_{2}a_{2}^{\dagger }a_{2}a_{1}^{+}\rangle ,
\nonumber
\end{eqnarray}%
\begin{eqnarray}
\frac{d\langle a_{1}a_{1}^{+}a_{1}a_{1}^{+}\rangle }{dt} &=&2(\alpha
_{1}+\alpha _{1}^{\ast }-2\kappa _{1})\langle
a_{1}a_{1}^{+}a_{1}a_{1}^{+}\rangle  \\
&&+(\alpha _{1}+\alpha _{1}^{\ast }+6\kappa _{1})\langle
a_{1}^{+}a_{1}\rangle   \nonumber \\
&&+2\alpha _{12}\langle a_{1}^{\dagger }a_{1}a_{1}^{+}a_{2}\rangle +2\alpha
_{12}^{\ast }\langle a_{1}a_{1}^{+}a_{1}a_{2}^{\dagger }\rangle   \nonumber
\\
&&+\alpha _{12}\langle a_{2}a_{1}^{+}\rangle +\alpha _{12}^{\ast }\langle
a_{1}a_{2}^{\dagger }\rangle   \nonumber \\
&&+4\kappa _{1}+\alpha _{1}+\alpha _{1}^{\ast },  \nonumber
\end{eqnarray}%
\begin{eqnarray}
\frac{d\langle a_{1}^{+}a_{1}a_{2}^{\dagger }a_{2}\rangle }{dt} &=&(\alpha
_{1}+\alpha _{1}^{\ast }+\alpha _{2}+\alpha _{2}^{\ast }-2\kappa
_{1}-2\kappa _{2})\langle a_{1}^{+}a_{1}a_{2}^{\dagger }a_{2}\rangle  
\nonumber \\
&&+(\alpha _{1}+\alpha _{1}^{\ast })\langle a_{2}^{+}a_{2}\rangle +(\alpha
_{2}+\alpha _{2}^{\ast })\langle a_{1}^{\dagger }a_{1}\rangle   \nonumber \\
&&+\alpha _{21}^{\ast }\langle a_{1}^{+}a_{1}a_{1}^{\dagger }a_{2}\rangle
+\alpha _{21}\langle a_{1}a_{1}^{+}a_{1}a_{2}^{\dagger }\rangle   \nonumber
\\
&&+\alpha _{12}\langle a_{2}a_{2}^{\dagger }a_{2}a_{1}^{+}\rangle \alpha
_{12}^{\ast }\langle a_{2}^{\dagger }a_{2}a_{2}^{+}a_{1}\rangle .
\end{eqnarray}%
Substituting the subscript 1 (2) with 2 (1) and then making their Hermitian
conjugate through (B1) to (B5), we can obtain the other seven differential
equations. The totall thirteen differenttial equations will be a closed set.
We can numerical solve it.

\end{document}